\documentclass{iau}
\usepackage{graphicx}

\title[JD 11.~~Distance to the SMC] %% give here short title %%
{Distance to the SMC from \\ eclipsing binaries. }

\author[Graczyk et al.]   %% give here short author list %%
{Dariusz Graczyk$^1$,
Grzegorz Pietrzy{\'n}ski$^{2,1}$, 
 Bogumi{\l} Pilecki$^{1,2}$,
Ian B. Thompson$^3$,
Wolfgang Gieren$^1$,
Piotr Konorski$^2$,
Andrzej Udalski$^2$ \and
Igor Soszy{\'n}ski$^2$}

\affiliation{$^1$ Departamento de Astronom{\'i}a, Universidad de Concepci{\'o}n, Casilla 160-C, Concepci{\'o}n, Chile \\ email: {\tt darek@astro-udec.cl, bpilecki@astro-udec.cl, wgieren@astro-udec.cl} \\[\affilskip]
$^2$ Obserwatorium Astronomiczne, Uniwersytet Warszawski, \\Al. Ujazdowskie 4, 00-478 Warszawa, Poland \\email: {\tt pietrzyn@astrouw.edu.pl, piokon@astrouw.edu.pl, udalski@astrouw.edu.pl\\soszynsk@astrouw.edu.pl} \\[\affilskip]
$^3$ Carnegie Observatories, 813 Santa Barbara Street, Pasadena, CA 911101-1292, USA\\
{\tt ian@obs.carnegiescience.edu} \\[\affilskip]
}

\pubyear{2013}
\volume{289}  %% insert here IAU Symposium No.
\jname{Advancing the physics of cosmic distances}
\editors{Richard de Grijs \& Giuseppe Bono, eds.}
\begin{document}

\maketitle

\begin{abstract}
The preliminary distance to a long period eclipsing binary in the Small Magellanic Cloud SMC108.1.14904 is presented. The binary system contains two non-active G-type bright giants. The orbital period is 185 days and the orbit is circular. Using surface brightness calibration we calculated distance modulus to the system (m-M)= 19.02 $\pm$ 0.04 (stat.) $\pm$ 0.05 (syst.) mag, where systematic error is dominated by uncertainty of surface brightness calibration. This is a second eclipsing binary in the SMC analysed by our team.    
\keywords{binaries: eclipsing, galaxies: distances and redshifts, galaxies: individual (SMC)}
\end{abstract}

\firstsection % if your document starts with a section,
\section{Introduction}

Determination of distances to the eclipsing binary stars is an important subproject of the ARAUCARIA international project devoted to set zero point calibration to some important standard candles (\cite[Pietrzy{\'n}ski \& Gieren 2002]{PietGie02}). One of our task is to precisely measure the distance to the Large Magellanic Cloud, and also its neighbour the Small Magellanic Cloud. In order to achieve this goal we selected 20 late type eclipsing binaries in the LMC and 8 similar systems in the SMC. Results regarding the LMC's eclipsing binaries are presented in a separate contribution (\cite[Pietrzy{\'n}ski et al. 2013]{Pietrzyn13}) to this symposium and we will focus on the SMC. 

\begin{figure}[b]
% \vspace*{-2.0 cm}
\begin{center}
 \includegraphics[width=3.3in]{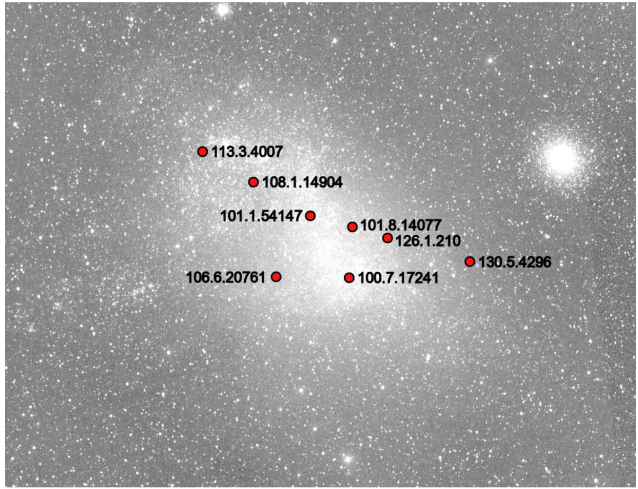} 
% \vspace*{-1.0 cm}
 \caption{8 late-type eclipsing binary stars selected in the SMC for distance measurements.}
   \label{fig1}
\end{center}
\end{figure}

Figure\,\ref{fig1} is presenting the position of our target  within the body of the SMC. All binaries contain photometrically non-variable (at least on the 0.01 mag level) G- or K-type giants with well detached configuration. One of these eclipsing binary stars was already analysed by us (\cite[Graczyk et al. 2012]{Graczyk12}). The main objective behind this subproject is an independent comparison of the distance to the LMC obtained from similar eclipsing binaries and a calibration of some distance indicators within lower metallicity regime. We briefly report here analysis of the second system SMC108.1.14904.

\section{Procedure to obtain the distance.}

To determine a distance to a particular eclipsing binary we employed the following procedure:
{\small
\begin{itemize}
\item I-band and V-band light curves are collected within OGLE project (\cite[Udalski et al. 1997]{Udalski97})
\item High-resolution spectra are taken to sample an orbit.
\item J-band and K-band infrared photometry is secured outside eclipses.
\item The Broadening Function (\cite[Rucinski 1992]{Rucinski92}) and TODCOR (\cite[Mazeh \& Zucker 1994]{MaZu94}) formalisms are used to obtain radial velocities of both components.
\item Extinction is estimated from \cite[Haschke et al. (2011)]{Haschke11} reddening maps.
\item From preliminary light curve solution and derredened individual $V-K$ colors approximate temperatures of the components are calculated.
\item Simultaneous analysis of radial velocities and light curves using Wilson-Devinney 2007 code (\cite[Wilson \& Devinney 1971]{WD71}, \cite[van Hamme \& Wilson 2007]{WD07}; hereafter WD) is undertaken.
\item Updated $V-K$ colors and absolute dimensions of the components are used to calculate a distance to the binary from empirical surface brightness calibration S$_{\rm V}$-($V-K$).
\item New temperatures are calculated according the updated ($V-K$) indices.
\item Repeat the last three points until full consistency of the model parameters is obtained. 
\end{itemize}
}

\section{Absolute dimensions and the distance to the system.}
\begin{figure}[b]
% \vspace*{-2.0 cm}
\begin{center}
 \includegraphics[width=3.6in]{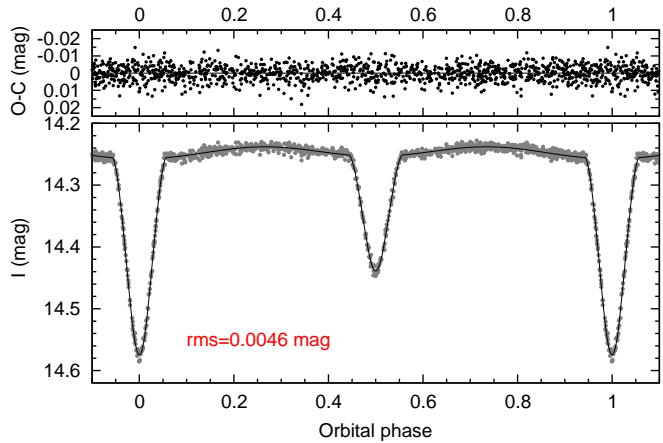} 
% \vspace*{-1.0 cm}
 \caption{The I-band light curve solution to the system SMC108.1.14904. The root mean square of the residuals is on the level of intrinsic photometric noise.}
   \label{fig2}
\end{center}
\end{figure}

From simultaneous analysis of radial velocities and light curves we derived a set of fundamental physical parameters of the system and its components. WD fits to both I-band and V-band light curves are excellent without any systematic deviations (see Fig.\,\ref{fig2}). From the solution we infer that both components are of equal mass with the primary star being smaller, hotter and less luminous of the two. It is worth to note that the primary lies just outside of the red border of the cepheid instability strip in the SMC and, in agreement with theoretical expectations, we could not detect any sign of larger pulsations in this star. 

  \begin{figure}[b]
% \vspace*{-2.0 cm}
\begin{center}
 \includegraphics[width=3.6in]{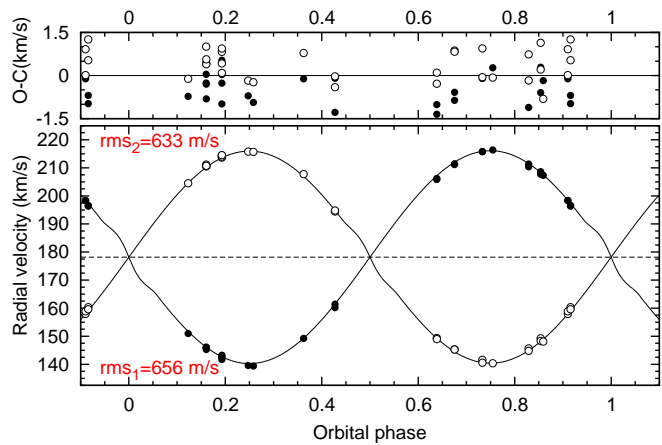} 
% \vspace*{-1.0 cm}
 \caption{The radial velocity solution to both components. The radial velocities of the primary (filled circles) are blueshifted by $-0.7$ km/s in respect to the secondary's velocities.}
   \label{fig3}
\end{center}
\end{figure}
\

\begin{table}
\begin{center}
\caption{The basic astrophysical parameters of the SMC108.1.14904 system.}
\label{tab1}
{\small
\begin{tabular}{|l|cc|}\hline
{\bf Parameter} & {\bf Primary Star} & {\bf Secondary Star} \\ \hline
Spectral Type & G2 II & G8 II \\
Distance Modulus [mag] &19.021 & 19.019\\ 
Orbital Period [$d$] & \multicolumn{2}{c|}{185.1}\\
Semimajor Axis [$R_\odot$] & \multicolumn{2}{c|}{282.7} \\
Orbital Inclination [deg] & \multicolumn{2}{c|}{78.8} \\
Systemic Velocity  [km/s] & 177.8 & 178.5\\
Radius [$R_\odot$] & 47.3 & 63.8 \\
Mass    [$M_\odot$] & 4.42 & 4.43 \\
Temperature $T_{eff}$  [$K$]  & 5470 & 4855 \\ 
Observed $V$ magnitude &15.93 & 15.99\\
Observed ($V-K$) color &1.85 & 2.39 \\ 
Extinction $E(B-V)$ & \multicolumn{2}{c|}{0.051}\\ \hline
\end{tabular}
}
\end{center}
\end{table}

\begin{figure}[b]
% \vspace*{-2.0 cm}
\begin{center}
 \includegraphics[width=3.7in]{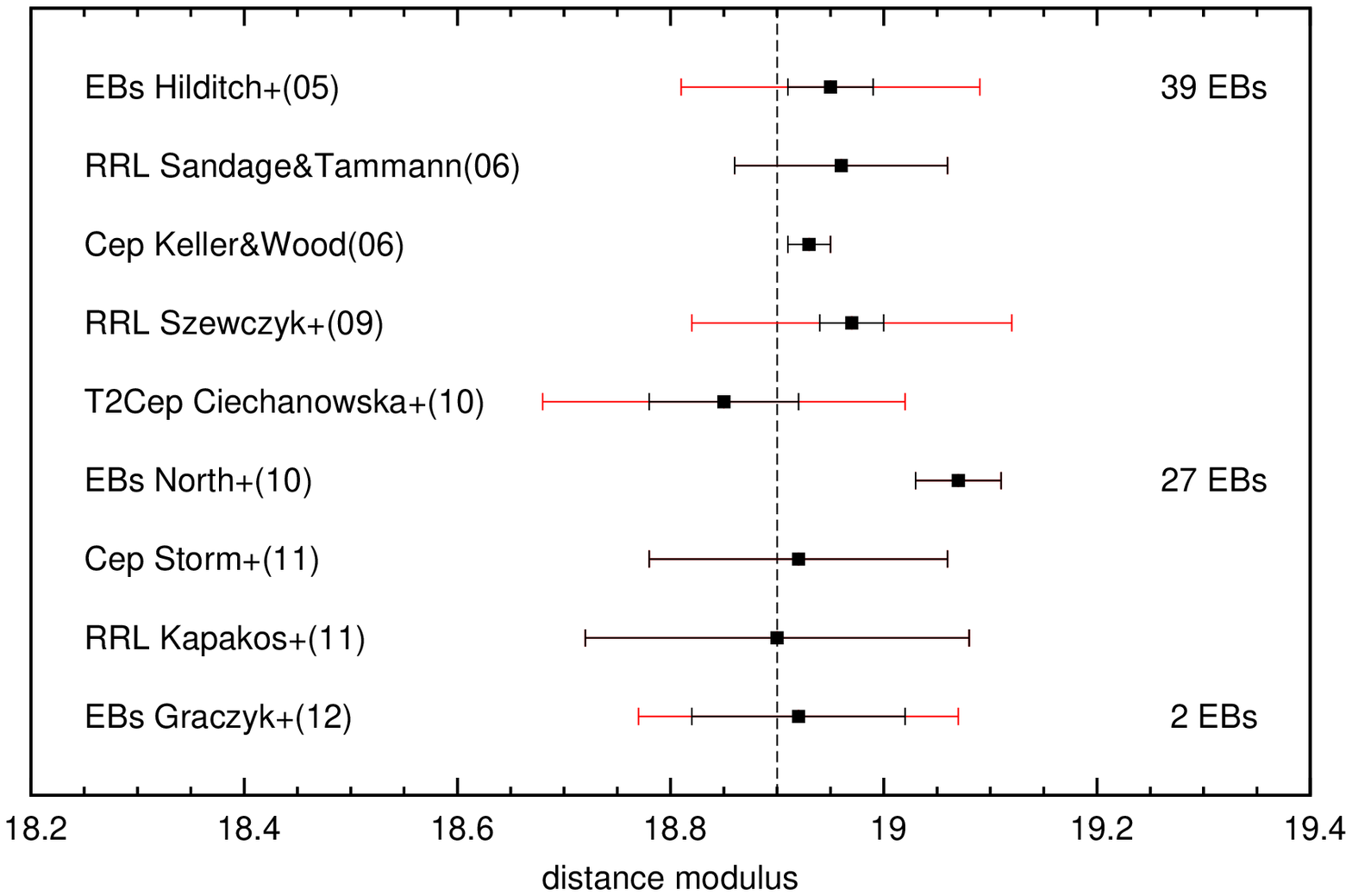} 
% \vspace*{-1.0 cm}
 \caption{Recent distance measurements to the SMC. The "standard" distance modulus to the SMC 18.90 mag is also highlighted. Internal error box denotes statistical error and external error box is systematic error. }
   \label{fig4}
\end{center}
\end{figure}

Table~\ref{tab1} presents astrophysical parameters of the system. The distance was derived using \cite[di Benedetto (2005)]{diB05} infrared calibration of V-band surface brightness relation. The resulting distance modulus is ($m-M$) = 19.02 $\pm$ 0.04 (statistical) $\pm$ 0.05 (systematic) mag. The primary source of statistical error is reddening uncertainty and systematic error is dominated by di Benedetto's relation uncertainty. The difference of the distance moduli of both components is very small which serves as an independent check of model consistency. The main line of an improvement of this distance determination would be better temperature and reddening estimation from analysis of disentangled spectra and also securing more infrared K-band photometry.

\section{Concluding remarks}
The distance modulus to another late eclipsing binary star SMC113.3.4007 was reported by \cite[Graczyk et al. (2012)]{Graczyk12} to be 18.83 mag. Combining both results we obtain mean distance modulus 18.92 mag. Both stars lie in north-east part of the SMC thus this result is not fully representative to the whole galaxy as there is a distance gradient across its visible disc, with north-east part being closer to us than the center.  Figure~\ref{fig4} gives summary of recent distance measurements to the SMC with different methods employed. After approximate subtraction of systematic uncertainties common to all methods (e.g. zero points of photometry, foreground reddening) the mean weighted distance modulus to the SMC from all determinations is 18.95 mag with the reduced $\chi^2=0.69$ what may suggest some psychological phenomena in clustering of the results.

%\begin{discussion}

%\end{discussion}


\begin{thebibliography}{}

\bibitem[di Benedetto (2005)]{diB05}{di Benedetto, G. P.}, 2005,
\textit{MNRAS}, 357, 174

\bibitem[Ciechanowska \etal\ (2010)]{Ciech10}{Ciechanowska, A., et al.}, 2010,
\textit{Acta Astron.}, 60, 233

\bibitem[Graczyk \etal\ (2012)]{Graczyk12}{Graczyk, D., et al.} 2012,
\textit{ApJ}, 750, 144

\bibitem[Haschke \etal\ (2011)]{Haschke11}{Haschke, R., Grebel, E. K., \& Duffau, S.}, 2011,
\textit{AJ}, 141, 158

\bibitem[Hilditch \etal\ (2005)]{Hild05}{Hilditch, R. W., Howarth, I. D., \& Harries, T. J.}, 2005,
\textit{MNRAS}, 357, 304

\bibitem[Kapakos \etal\ (2011)]{Kapak11}{Kapakos, E., Hatzidimitriou, D., \& Soszy{\'n}ski, I.}, 2011,
\textit{MNRAS}, 415, 1366

\bibitem[Keller \& Wood (2006)]{KeWo06}{Keller, S. C., \& Wood, P. R.}, 2006,
\textit{ApJ}, 642, 834

\bibitem[Mazeh \& Zucker (1994)]{MaZu94}{Mazeh, T., \& Zucker, S.} 1994,
\textit{Ap{\rm\&}SS}, 212, 349

\bibitem[North \etal\ (2010)]{Nort10}{North, P., Gauderon, R., Barblan, F., \& Royer, F.}, 2010,
\textit{A{\rm\&}A}, 520, 74 

\bibitem[Pietrzy{\'n}ski \& Gieren (2002)]{PietGie02}{Pietrzy{\'n}ski, G., \& Gieren, W.} 2002,
\textit{AJ}, 124, 2633

\bibitem[Pietrzy{\'n}ski \etal\ (2013)]{Pietrzyn13}{Pietrzy{\'n}ski, G., et al.} 2013, 
\textit{this conference proceedings}

\bibitem[Rucinski (1992)]{Rucinski92}{Rucinski, S. M.}, 1992,
\textit{AJ}, 104, 1968

\bibitem[Sandage \& Tammann (2006)]{SaTa06}{Sandage, A., \& Tammanna, G. A.}, 2006,
\textit{ARA{\rm\&}A}, 44, 93

\bibitem[Storm \etal\ (2011)]{Storm11}{Storm, J., et al.}, 2011,
\textit{A{\rm\&}A}, 534, 95

\bibitem[Szewczyk (2009)]{Szewc09}{Szewczyk, O., et al.}, 2009,
\textit{AJ}, 138,1661

\bibitem[Udalski \etal\ (1997)]{Udalsi97}{Udalski, A., Kubiak, M., \& Szyma{\'n}ski, M.}, 1997,
\textit{Acta Astron.}, 47, 319 

\bibitem[van Hamme \& Wilson (2007)]{WD07}{van Hamme, W., \& Wilson, R. E.}, 2007,
\textit{ApJ}, 661, 1129

\bibitem[Wilson \& Devinney (1971)]{WD71}{Wilson, R. E., \& Devinney, E. J.}, 1971,
\textit{ApJ}, 166, 605

\end{thebibliography}
\end{document}